# Sb doping effect on transport behavior in the topological insulator $Bi_2Se_3$

Shu-Wei Wang[1,2,*], Hang Chi[3], and Jui-Che Chung[1,4]

[1]*Institute of Microelectronics, Department of Electrical Engineering, National Cheng Kung University, Tainan 701, Taiwan*
[2]*Academy of Innovative Semiconductor and Sustainable Manufacturing, National Cheng Kung University, Tainan 701, Taiwan*
[3]*Department of Physics, University of Ottawa, Ottawa, Ontario K1N 6N5, Canada*
[4]*Department of Physics, National Sun Yat-sen University, Kaohsiung 804, Taiwan*

[*]Corresponding author: shuweiwang@gs.ncku.edu.tw

**Abstract**

Bismuth selenide ($Bi_2Se_3$) is a good topological insulator (TI) with its surface band Dirac point inside the bulk bandgap. However, $Bi_2Se_3$ films grown by molecular beam epitaxy (MBE) often require tuning of the Fermi level near the Dirac point for optimal proximity effect with magnetic or superconducting materials. In this study, we achieve the control of the Fermi level in MBE-grown $Bi_2Se_3$ thin films by antimony (Sb) doping and systematically investigate the transport properties of these $Bi_{2-x}Sb_xSe_3$ films with different doping concentrations. Excellent topological surface conduction is attained, and weak antilocalization is observed in all Sb-doped $Bi_2Se_3$ films. While the carrier mobility shows no dependence on the Sb concentrations, indicating that the phonon scattering dominates over the impurity scattering from Sb dopants, the coherence length varies significantly with the Sb doping level at low temperatures (< 30 K), highlighting the non-negligible electron-electron interactions in the low temperature regime. Furthermore, EuS/$Bi_{2-x}Sb_xSe_3$ heterostructures are fabricated to explore proximity-induced ferromagnetism in the TI surface states. However, the long-range magnetic order is not formed in the TI surface states under our growth conditions. Our results emphasize the critical role of interface quality for realizing exchange coupling. This work offers new insights into the interplay of disorder, decoherence, and scattering mechanisms in Sb-doped $Bi_2Se_3$ thin films, providing guidance for the future study of the proximity effect in heterostructures involving Sb-doped $Bi_2Se_3$.

After the realization of the quantum spin Hall effect in the HgTe/CdTe quantum well in 2007 [1], people tried to look for the three-dimensional (3D) analog of this time-reversal invariant system. The 3D time-reversal invariant system with $Z_2$ topological order is called a topological insulator (TI) owing to its topologically non-trivial band structure [2–5]. In 2008, the topological surface states were observed in $Bi_{1-x}Sb_x$ alloys using angle-resolved photoemission spectroscopy (ARPES) [6], marking the discovery of the first 3D TI. However, $Bi_{1-x}Sb_x$ is not an ideal system for further study of TIs due to its complicated band structure and small bandgap [6]. This motivated the search for other 3D TIs with simpler band structures. Later, it was predicted that $Bi_2Se_3$, $Bi_2Te_3$, and $Sb_2Te_3$ are potential candidates for 3D TIs due to the strong spin-orbit coupling inside these materials [7]. The size of bandgaps of these materials is larger and their band structures are simpler compared to that of the $Bi_{1-x}Sb_x$ alloy. Shortly after the theoretical prediction, these materials were experimentally verified to be 3D TIs and were regarded as second-generation 3D TIs [8–13]. Among the second-generation 3D TIs, $Bi_2Se_3$ attracts particular attention. The band structure of $Bi_2Se_3$ measured by ARPES shows a highly symmetric single Dirac cone with a bulk bandgap size of 0.3 eV, which implies that the topological surface states can survive at room temperature [10,14]. The large bulk bandgap and the simple band structure suggest that $Bi_2Se_3$ not only is suitable for the fundamental study of TIs, but also has great potential in practical applications.

In the early stage, the studies of the TI surface states were mostly performed by ARPES because ARPES is able to distinguish the bulk band and surface band unambiguously. In contrast, the investigation of the properties of surface states of 3D TIs by transport measurements is usually difficult due to the mixed bulk and surface contributions. In order to clearly study the properties of the two-dimensional (2D) topological surface states of 3D TIs, the influence of the bulk states should be minimized. In practice, this is typically attained by tuning the position of the Fermi level away from the bulk conduction band (BCB) and bulk valence band (BVB) edges and bringing it as close as possible to the Dirac point of the surface band. Several approaches have been proposed to achieve this goal, including doping complementary elements [10,12–14], electric gating [15,16], and growing ternary or quaternary compounds of $Bi_{2-x}Sb_xSe_{3-y}Te_y$ (where $0 \leq x \leq 2$ and $0 \leq y \leq 3$) [17–20]. Making $Bi_{2-x}Sb_xSe_{3-y}Te_y$ compound is particularly useful because it allows tuning of the Fermi level, the position of the Dirac point, and the size of the bulk gap, giving rise to TI systems with significantly less bulk conduction and much more pronounced surface signals in transport measurements.

In this work, we tune the Fermi level of $Bi_2Se_3$ thin films by doping Sb atoms to replace Bi atoms and form $Bi_{2-x}Sb_xSe_3$ compounds during molecular beam epitaxy (MBE), with the attempt to move the Fermi level into the magnetic gap opened in the surface band by the adjacent EuS layer. If the ferromagnetic phase can form in the TI surface states by the exchange coupling from the EuS layer via the proximity effect, the EuS/TI heterostructure can provide a platform for studying exotic phenomena in magnetic TIs, such as the quantum anomalous Hall effect [21], the axion electrodynamics [22,23], and the topological magnetoelectric effect [24–28] at higher temperatures. While Sb-doped $Bi_2Se_3$ systems have been explored in previous studies, most of them focused on tuning the Fermi level to isolate surface states or on band structure characterization via ARPES [17,19,20,29]. The impact of Sb doping on the underlying scattering and decoherence mechanisms in MBE-grown $Bi_2Se_3$ thin films remains underexplored. These mechanisms are crucial for understanding the limits of carrier mobility and quantum coherence in topological transport. Therefore, this work addresses this gap by providing a systematic transport study of $Bi_{2-x}Sb_xSe_3$ films with varying Sb concentrations, offering new insights into how disorder and decoherence evolve with doping in epitaxially grown topological insulator thin films. We observed weak antilocalization (WAL) in all our samples with different Sb concentrations, which implies that surface conduction dominates. However, under the sample growth conditions explored within the scope of this report, the EuS layer was unable to induce long-range ferromagnetic order in the TI surface states at the interface. Notably, we found that the $Bi_{2-x}Sb_xSe_3$ films with different Sb doping concentrations showed similar carrier mobility values, suggesting that the impurity scattering from the Sb dopants is not the main reason affecting the mobility. Nevertheless, at sufficiently low temperatures, the coherence length decreased when the Sb doping level increased, indicating that the decoherence in the quantum diffusive transport regime is enhanced by the doped impurities.

The $Bi_{2-x}Sb_xSe_3$ films used in this study were grown on 0.5 mm thick heat-treated $SrTiO_3$ (111) substrates by MBE in a homemade MBE chamber. The nominal Sb doping levels were chosen to be $x = 0.28$ and $x = 0.72$. After the growth of a 6 nm $Bi_{2-x}Sb_xSe_3$ layer on the substrate, a 3-nm-thick EuS layer was grown *in situ*. Finally, a 2-nm-thick aluminum oxide capping layer was deposited on the top to protect the sample from degradation upon air exposure. The heterostructure was then etched into a Hall bar shape by standard e-beam lithography. Cr/Au electrodes were thermally evaporated to form Ohmic contacts of the Hall bar. Finally, the Hall bar device was electrically connected to a sample holder by gold wires using a wire bonder. The electrical measurements were carried out in a Quantum Design DynaCool Physical Property Measurement System (PPMS) with a base temperature of 2 K and a maximum magnetic field of 9 T.

The magnetic field ($\mu_0H$) dependence of the longitudinal resistance ($R_{xx}$) at different temperatures ($T$) is shown in the upper panels of Figs. 1(a) and 1(b). The WAL feature can be observed in both the $Bi_{1.72}Sb_{0.28}Se_3$ and $Bi_{1.28}Sb_{0.72}Se_3$ films from 2 K to 60 K. In this temperature range, the carrier concentrations obtained by the Hall effect are $\sim 5 \times 10^{13}$ cm$^{-2}$ and $\sim 1 \times 10^{13}$ cm$^{-2}$ for the $Bi_{1.72}Sb_{0.28}Se_3$ and $Bi_{1.28}Sb_{0.72}Se_3$ films, respectively (see more information in the supplementary material [30]). Both the $Bi_{1.72}Sb_{0.28}Se_3$ and $Bi_{1.28}Sb_{0.72}Se_3$ films show *n*-type transport. The order of carrier concentrations and the carrier type in our Bi$_{2-x}$Sb$_x$Se$_3$ samples agree well with the prior study regarding Sb-doped Bi$_2$Se$_3$ [29]. In the lower panels of Figs. 1(a) and 1(b), the ordinate is displayed in normalized magnetoresistance $R_{xx}/R_{xx}(0)$, where $R_{xx}(0)$ is the $R_{xx}$ at zero magnetic field, i.e., the vertex of the WAL dip. The WAL dips at zero magnetic field become less pronounced as the temperature increases, possibly due to the weakened quantum coherence at higher temperatures. Figure 2 shows the temperature dependence of the carrier mobility $\mu$ and $R_{xx}$ at zero magnetic field, $R_{xx}(0)$. The carrier mobility decreases with increasing temperature, implying that the main scattering mechanism is phonon scattering from the lattice instead of ionized impurity scattering from the Sb dopants [31]. In the case where ionized impurity scattering dominates, the average carrier velocity increases with increasing temperature, and the charge carriers have a higher energy to overcome the hindrance from the ionized impurity. Hence, the carrier mobility increases as the temperature rises. On the other hand, when phonon scattering dominates, the vibration of lattice is enhanced at higher temperatures, which would impede the propagation of charge carriers and lead to decreasing mobility as the temperature increases. This phonon-dominated scattering mechanism is also evidenced by the similar mobility values in the Bi$_{1.72}$Sb$_{0.28}$Se$_3$ and Bi$_{1.28}$Sb$_{0.72}$Se$_3$ films and the $T^{-3/2}$ dependence of $\mu$ [31](see the discussion for Fig. S7 in the supplementary material [30]), suggesting that the Sb doping level is not the main factor affecting the carrier mobility in Bi$_{2-x}$Sb$_x$Se$_3$ films. Contrary to the temperature dependence of the carrier mobility, the zero-field $R_{xx}$ increases with increasing temperature, which is a metallic temperature dependence, suggesting that the topological surface states are dominant in our transport measurements, and the bulk states of TIs are negligible [30,32]. The dominant surface transport can also be corroborated by the observed WAL behavior because the bulk states would give rise to weak localization (WL) instead of WAL [33].

For the detailed analysis of the WAL behavior, we employ the Hikami-Larkin-Nagaoka (HLN) formalism [34,35]:

$$\Delta\sigma_{xx}(B) = \sigma_{xx}(B) - \sigma_{xx}(0) = \frac{\alpha e^2}{\pi h}[\psi\left(\frac{1}{2} + \frac{B_\phi}{B}\right) - \ln(\frac{B_\phi}{B})]$$

where $\sigma_{xx}$ is the 2D longitudinal magnetoconductance, $\psi$ is the digamma function, $e$ is the elementary charge, $h$ is Planck's constant, $B_\phi$ is the dephasing field, $B$ equals to $\mu_0 H$ in our experiment, and $\alpha$ is a pre-factor that indicates the type of localization. The HLN equation describes the relation between $\Delta\sigma_{xx}$ and the magnetic field in the WAL regime and can be used for fitting to obtain the fitting parameters $\alpha$ and $B_\phi$. Figure 3 shows the fitting of the magnetoconductance data at various temperatures between -1 T and 1 T, in which the magnetic field is low enough to let the electronic system stay in the WAL regime. The fitting curves agree well with the experimental data. We can calculate the phase coherence length $l_\phi$ at each temperature by the relation $B_\phi = \hbar/4el_\phi^2$ [36–38]. The temperature dependence of $l_\phi$ and $\alpha$ is shown in Fig. 4. The phase coherence length decreases with increasing temperature, which is consistent with the observation in Fig. 1 that the WAL dip gradually shrinks as the temperature increases. Moreover, the phase coherence length in the $Bi_{1.72}Sb_{0.28}Se_3$ film is significantly larger than that in the $Bi_{1.28}Sb_{0.72}Se_3$ film below 30 K but becomes comparable over 30 K (see more statistical analysis in the supplementary material [30]). This phenomenon likely stems from the different predominant dephasing mechanisms in different temperature regimes [39–41]. Besides temperature (phonon), electron-electron interaction is another factor that affects the phase coherence length [33,37,42,43]. Below 30 K, the electron-electron interaction caused by the doped impurities is not negligible, so the decoherence mechanism is dominated by the electron-electron interaction [41]. As a result, the phase coherence length is reduced more in the sample with heavier Sb doping and stronger impurity effect, i.e., the $Bi_{1.28}Sb_{0.72}Se_3$ film, because the higher Sb content will introduce more disorder and enhance the electron-electron interaction. When the temperature exceeds 30 K, the electron-electron interaction becomes less important and phonon scattering takes over the main role of the dephasing mechanism. Hence the phase coherence lengths in the two samples become similar above 30 K. The $\alpha$ values at all temperatures are negative and vary from ~ -0.1 to ~ -0.4 for both $Bi_{2-x}Sb_xSe_3$ films, which is in good agreement with prior studies on WAL behavior in $Bi_2Se_3$ [33,35,44].

Next, we discuss the influence of different Sb doping levels on the band structures of the two $Bi_{2-x}Sb_xSe_3$ films and the possible reason why the EuS layer does not magnetically couple to the TI layer. Since both films exhibit *n*-type transport behavior in the Hall measurements, we can deduce that the Fermi levels cross the conduction band of the topological surface states. The observed WAL suggests that the Fermi level is well inside the bulk bandgap for both $Bi_{1.72}Sb_{0.28}Se_3$ and $Bi_{1.28}Sb_{0.72}Se_3$ films. The carrier concentration in the $Bi_{1.28}Sb_{0.72}Se_3$ film is significantly lower than that in the $Bi_{1.72}Sb_{0.28}Se_3$ film (see more data in the supplementary material [30]), which means that the doping of Sb atoms successfully changes the position of the Fermi level and move it closer to the Dirac point. The inferred band structure of $Bi_{1.72}Sb_{0.28}Se_3$ and $Bi_{1.28}Sb_{0.72}Se_3$ films are shown in Fig. S9. We note that the magnetic gap is not opened in the topological surface bands in both of our $Bi_{2-x}Sb_xSe_3$ films. Unlike the reports in Refs. [45–47], in which ferromagnetism is induced in the undoped $Bi_2Se_3$ films by the adjacent EuS layer through the proximity effect, the long-range ferromagnetic order is not formed in the TI surface states of our current $Bi_{2-x}Sb_xSe_3$ samples. EuS is an insulating ferromagnet with a Curie temperature ($T_C$) of ~ 15.7 K [45]. If the interfacial exchange coupling between the EuS and the $Bi_{2-x}Sb_xSe_3$ layers is strong enough, WL should be observed below $T_C$, and WAL should be recovered above $T_C$ [45]. In our samples, WAL is observed from 2 K to 60 K, indicating that the EuS layer and the $Bi_{2-x}Sb_xSe_3$ layer are not exchange coupled. This may be explained by the roughness and intermixing at the interface between the EuS layer and the $Bi_{2-x}Sb_xSe_3$ layer [45–47]. The lack of an atomically clean interface can tremendously weaken the proximity effect and undermine the coupling between the wavefunctions of the TI surface states and the spins in the EuS layer, hindering the emergence of the ferromagnetic phase in the TI films. More work is required to improve the interfacial coupling between the EuS layer and the epitaxial Bi$_{2-x}$Sb$_x$Se$_3$ layer, which may be achieved by optimizing the MBE processes favoring exquisitely clean interfaces, such as annealing the TI surface and removing residual Se from the growth environment before the deposition of EuS.

In summary, we systematically investigate the impact of Sb doping on the transport properties of MBE-grown Bi$_2$Se$_3$ thin films. Our results demonstrate that Sb substitution effectively tunes the Fermi level of the topological insulator material Bi$_2$Se$_3$. The resulting Bi$_{2-x}$Sb$_x$Se$_3$ thin films show excellent topological surface transport and negligible bulk contributions, as evidenced by the pronounced weak antilocalization observed across all doping levels. The weak antilocalization features weaken as the temperature increases, consistent with the reduced phase coherence lengths at elevated temperatures. Below 30 K, the phase coherence length is smaller in the Bi$_{2-x}$Sb$_x$Se$_3$ sample with a higher Sb doping level, which can be ascribed to the stronger electron-electron interactions at the higher Sb doping level. In contrast, above 30 K, the dephasing mechanism is dominated by thermal phonons, rendering the phase coherence length less sensitive to the Sb concentration. Importantly, the carrier mobility shows no dependence on the Sb doping concentration, implying that the impurity scattering from Sb dopants is not the primary factor affecting the carrier mobility in Bi$_{2-x}$Sb$_x$Se$_3$ films. We also explored magnetic proximity effects by interfacing Bi$_{2-x}$Sb$_x$Se$_3$ films with a EuS ferromagnetic insulator layer. However, no evidence of exchange-induced time-reversal symmetry breaking was observed, likely due to interfacial imperfections under the current growth conditions. These findings highlight the critical role of interface engineering in realizing magnetic topological phases in EuS/Bi$_{2-x}$Sb$_x$Se$_3$ heterostructures. Our study not only provides new insights into scattering and decoherence mechanisms in Sb-doped Bi$_2$Se$_3$ thin films but also establishes a foundation for future development of magnetic TI heterostructures with controlled disorder and optimized interfaces.

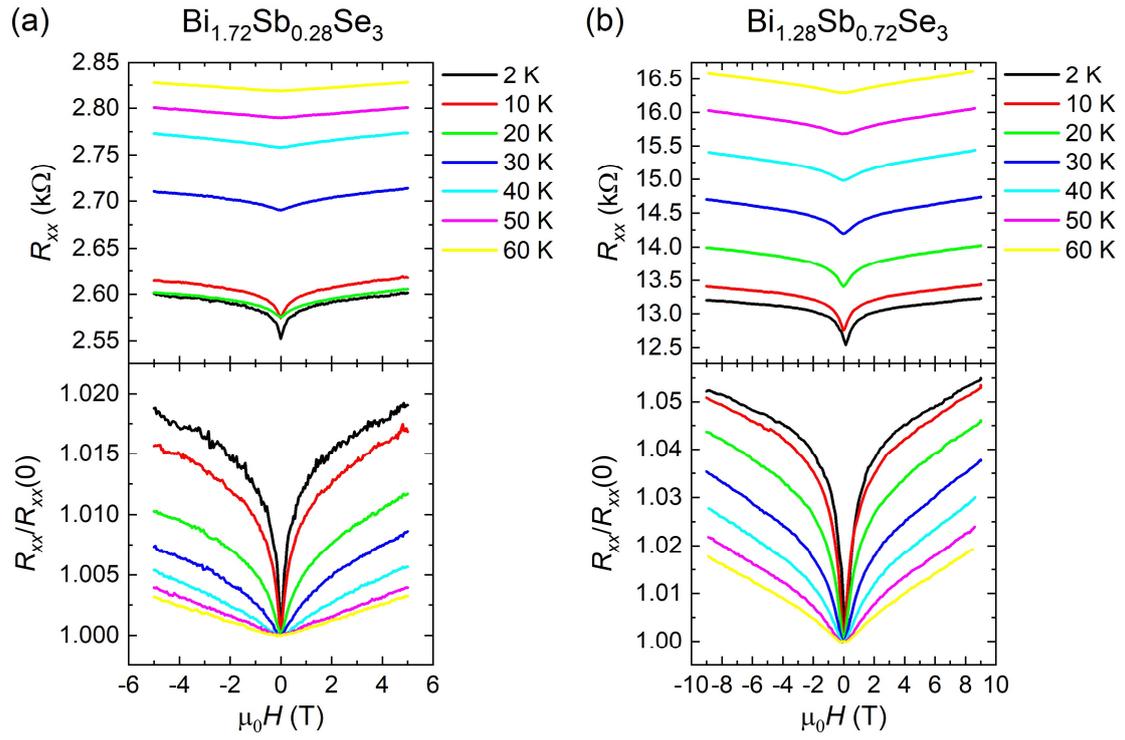

**FIG. 1.** WAL in $Bi_{2-x}Sb_xSe_3$ films. (a) The upper panel shows the magnetic field dependence of the longitudinal resistance in the $Bi_{1.72}Sb_{0.28}Se_3$ film from 2 K to 60 K. The lower panel shows the magnetic field dependence of the normalized magnetoresistance of the same film from 2 K to 60 K. (b) The upper panel shows the magnetic field dependence of the longitudinal resistance in the $Bi_{1.28}Sb_{0.72}Se_3$ film from 2 K to 60 K. The lower panel shows the magnetic field dependence of the normalized magnetoresistance of the same film from 2 K to 60 K.

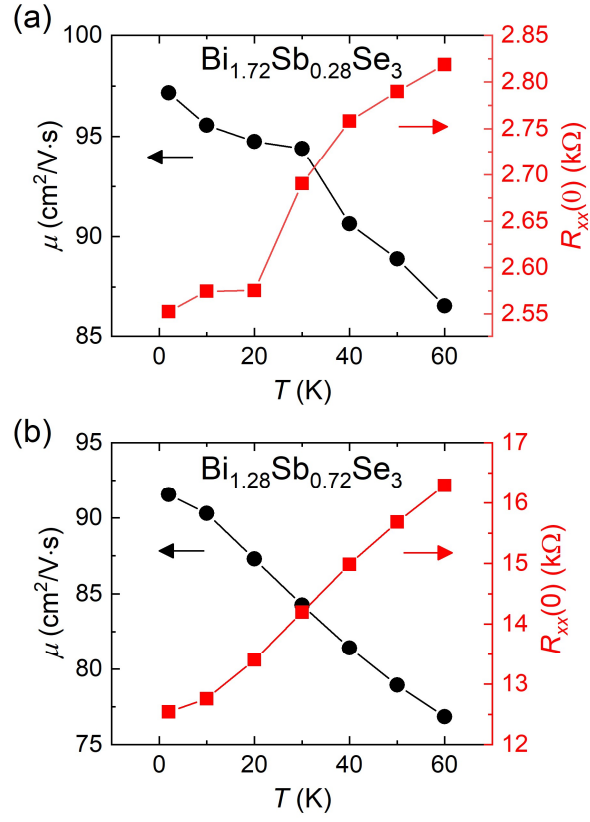

**FIG. 2.** Carrier mobility and zero-field longitudinal resistance as a function of temperature. (a) $T$ dependence of $\mu$ (black dots) and $R_{xx}(0)$ (red squares) of the $Bi_{1.72}Sb_{0.28}Se_3$ film. (b) $T$ dependence of $\mu$ (black dots) and $R_{xx}(0)$ (red squares) of the $Bi_{1.28}Sb_{0.72}Se_3$ film.

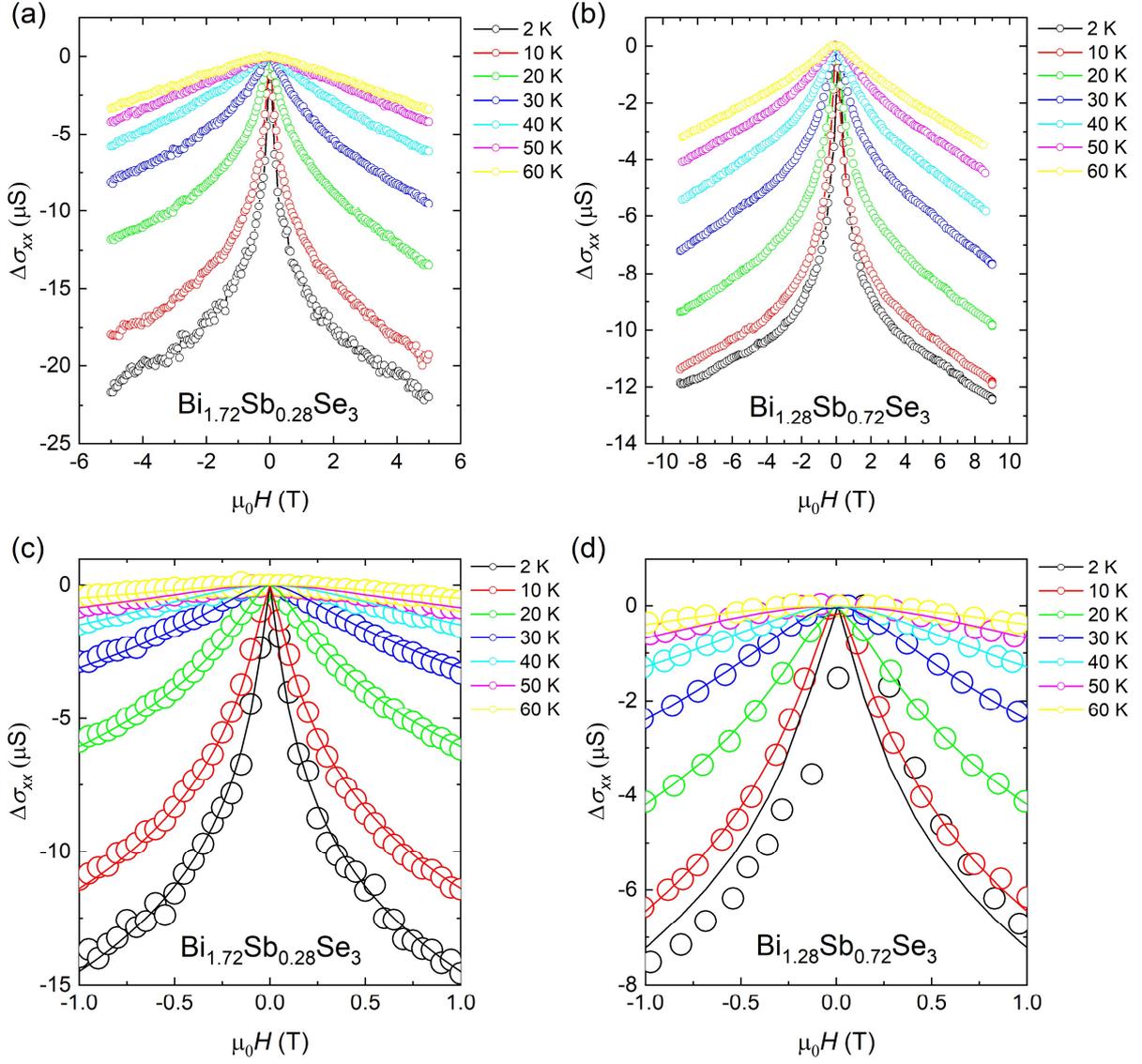

**FIG. 3.** 2D magnetoconductance and HLN fitting. (a) Magnetic field dependence of $\Delta\sigma_{xx}$ of the $Bi_{1.72}Sb_{0.28}Se_3$ film. (b) Magnetic field dependence of $\Delta\sigma_{xx}$ of the $Bi_{1.28}Sb_{0.72}Se_3$ film. (c) Zoomed-in view of (a) and the HLN fitting between -1 T and 1 T. (d) Zoomed-in view of (b) and the HLN fitting between -1 T and 1 T.

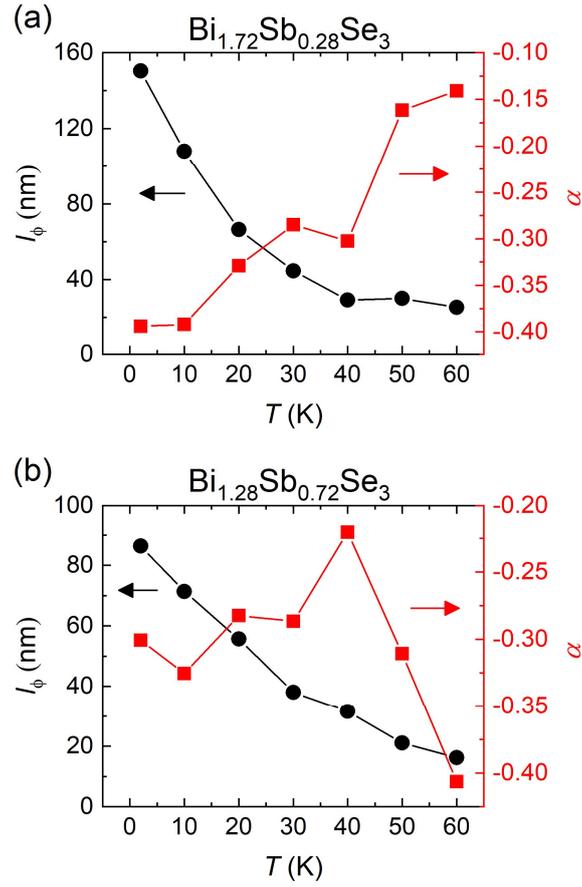

**FIG. 4.** Phase coherence length and HLN fitting pre-factor as a function of temperature. (a) $T$ dependence of $l_\phi$ (black dots) and $\alpha$ (red squares) of the $Bi_{1.72}Sb_{0.28}Se_3$ film. (b) $T$ dependence of $l_\phi$ (black dots) and $\alpha$ (red squares) of the $Bi_{1.28}Sb_{0.72}Se_3$ film.

**Supplementary Material**

The supplementary material contains further information about the sample growth, device dimensions, measurement setup, more detailed analyses of magnetotransport measurements, and other supporting data.

**Acknowledgements**

The authors thank Jagadeesh S. Moodera for the helpful discussions. S.-W. W. acknowledges the support from the National Science and Technology Council (Grant No. NSTC 112-2112-M-006-038-MY3) in Taiwan. H. C. acknowledges the support of the Army Research Office (W911NF-25-1-0215), the Canada Research Chairs (CRC) Program and the Natural Sciences and Engineering Research Council of Canada (NSERC), Discovery Grant Nos. RGPIN-2024-06497 and ALLRP 592642-2024. The work at MIT was supported by Army Research Office (Grant Nos. W911NF-20-2-0061 and DURIP W911NF-20-1-0074) and the National Science Foundation (Grant Nos. NSF-DMR 1700137 and NSF-DMR 1231319).

**Author Contributions**

S.-W. W. conceived and designed the experiments. H. C. conducted the MBE growth of the materials. S.-W. W. fabricated the devices and carried out the measurements. S.-W. W. and J.-C. Chung analyzed the data. S.-W. W. wrote the paper, with input from co-authors. All authors contributed to the scientific discussions.

**Conflict of Interest**

The authors have no conflicts to disclose.

**Data availability statement**

The data that support the findings of this study are available from the corresponding author upon reasonable request.

# References

[1] M. König, S. Wiedmann, C. Brüne, A. Roth, H. Buhmann, L. W. Molenkamp, X.-L. Qi, and S.-C. Zhang, Quantum spin Hall insulator state in HgTe quantum wells, Science **318**, 766 (2007).

[2] L. Fu, C. L. Kane, and E. J. Mele, Topological insulators in three dimensions, Phys. Rev. Lett. **98**, 106803 (2007).

[3] J. E. Moore and L. Balents, Topological invariants of time-reversal-invariant band structures, Phys. Rev. B **75**, 121306(R) (2007).

[4] R. Roy, $Z_2$ classification of quantum spin Hall systems: An approach using time-reversal invariance, Phys. Rev. B **79**, 195321 (2009).

[5] R. Roy, Topological phases and the quantum spin Hall effect in three dimensions, Phys. Rev. B **79**, 195322 (2009).

[6] D. Hsieh, D. Qian, L. Wray, Y. Xia, Y. S. Hor, R. J. Cava, and M. Z. Hasan, A topological Dirac insulator in a quantum spin Hall phase, Nature **452**, 970 (2008).

[7] H. Zhang, C.-X. Liu, X.-L. Qi, X. Dai, Z. Fang, and S.-C. Zhang, Topological insulators in $Bi_2Se_3$, $Bi_2Te_3$ and $Sb_2Te_3$ with a single Dirac cone on the surface, Nat. Phys. **5**, 438 (2009).

[8] Y. Xia, D. Qian, D. Hsieh, L. Wray, A. Pal, H. Lin, A. Bansil, D. Grauer, Y. S. Hor, R. J. Cava, and M. Z. Hasan, Observation of a large-gap topological-insulator class with a single Dirac cone on the surface, Nat. Phys. **5**, 398 (2009).

[9] Y. S. Hor, A. Richardella, P. Roushan, Y. Xia, J. G. Checkelsky, A. Yazdani, M. Z. Hasan, N. P. Ong, and R. J. Cava, *p*-type $Bi_2Se_3$ for topological insulator and low-temperature thermoelectric applications, Phys. Rev. B **79**, 195208 (2009).

[10] D. Hsieh, Y. Xia, D. Qian, L. Wray, J. H. Dil, F. Meier, J. Osterwalder, L. Patthey, J. G. Checkelsky, N. P. Ong, A. V. Fedorov, H. Lin, A. Bansil, D. Grauer, Y. S. Hor, R. J. Cava, and M. Z. Hasan, A tunable topological insulator in the spin helical Dirac transport regime, Nature **460**, 1101 (2009).

[11] S. R. Park, W. S. Jung, Chul Kim, D. J. Song, C. Kim, S. Kimura, K. D. Lee, and N. Hur, Quasiparticle scattering and the protected nature of the topological states in a parent topological insulator $Bi_2Se_3$, Phys. Rev. B **81**, 041405 (2010).

[12] D. Hsieh, Y. Xia, D. Qian, L. Wray, F. Meier, J. H. Dil, J. Osterwalder, L. Patthey, A. V. Fedorov, H. Lin, A. Bansil, D. Grauer, Y. S. Hor, R. J. Cava, and M. Z. Hasan, Observation of time-reversal-protected single-Dirac-cone topological-insulator states in $Bi_2Te_3$ and $Sb_2Te_3$, Phys. Rev. Lett. **103**, 146401 (2009).

[13] Y. L. Chen, J. G. Analytis, J.-H. Chu, Z. K. Liu, S.-K. Mo, X. L. Qi, H. J. Zhang, D. H. Lu, X. Dai, Z. Fang, S. C. Zhang, I. R. Fisher, Z. Hussain, and Z.-X. Shen, Experimental realization of a three-dimensional topological insulator, $Bi_2Te_3$, Science **325**, 178 (2009).

[14] Y. L. Chen, J.-H. Chu, J. G. Analytis, Z. K. Liu, K. Igarashi, H.-H. Kuo, X. L. Qi, S. K. Mo, R. G. Moore, D. H. Lu, M. Hashimoto, T. Sasagawa, S. C. Zhang, I. R. Fisher, Z. Hussain, and Z. X. Shen, Massive Dirac fermion on the surface of a magnetically doped topological insulator, Science **329**, 659 (2010).

[15] B. Sacépé, J. B. Oostinga, J. Li, A. Ubaldini, N. J. G. Couto, E. Giannini, and A. F. Morpurgo, Gate-tuned normal and superconducting transport at the surface of a topological insulator, Nat. Commun. **2**, 575 (2011).

[16] J. G. Checkelsky, Y. S. Hor, R. J. Cava, and N. P. Ong, Bulk band gap and surface state conduction observed in voltage-tuned crystals of the topological insulator $Bi_2Se_3$, Phys. Rev. Lett. **106**, 196801 (2011).

[17] J. Zhang, C.-Z. Chang, Z. Zhang, J. Wen, X. Feng, K. Li, M. Liu, K. He, L. Wang, X. Chen, Q.-K. Xue, X. Ma and Y. Wang, Band structure engineering in $(Bi_{1-x}Sb_x)_2Te_3$ ternary topological insulators, Nat. Commun. **2**, 574 (2011).

[18] Z. Ren, A. A. Taskin, S. Sasaki, K. Segawa, and Y. Ando, Optimizing $Bi_{2-x}Sb_xTe_{3-y}Se_y$ solid solutions to approach the intrinsic topological insulator regime, Phys. Rev. B **84**, 165311 (2011).

[19] C. Niu, Y. Dai, Y. Zhu, Y. Ma, L. Yu, S. Han and B. Huang, Realization of tunable Dirac cone and insulating bulk states in topological insulators $(Bi_{1-x}Sb_x)_2Te_3$, Sci. Rep. **2**, 976 (2012).

[20] C. H. Lee, R. He, Z. Wang, R. L. J. Qiu, A. Kumar, C. Delaney, B. Beck, T. E. Kidd, C. C. Chancey, R. M. Sankaran and X. P. A. Gao, Metal–insulator transition in variably doped $(Bi_{1-x}Sb_x)_2Se_3$ nanosheets, Nanoscale, **5**, 4337 (2013).

[21] C.-Z. Chang, C.-X. Liu, and A. H. MacDonald, Colloquium: Quantum anomalous Hall effect, Rev. Mod. Phys. **95**, 011002 (2023).

[22] M. Mogi, M. Kawamura, A. Tsukazaki, R. Yoshimi, K. S. Takahashi, M. Kawasaki, and Y. Tokura, Tailoring tricolor structure of magnetic topological insulator for robust axion insulator, Science Advances **3**, eaao1669 (2017).

[23] D. Xiao, J. Jiang, J.-H. Shin, W. Wang, F. Wang, Y.-F. Zhao, C. Liu, W. Wu, M. H. W. Chan, N. Samarth, and C.-Z. Chang, Realization of the axion insulator state in quantum anomalous Hall sandwich heterostructures, Phys. Rev. Lett. **120**, 056801 (2018).

[24] X.-L. Qi, T. L. Hughes, and S.-C. Zhang, Topological field theory of time-reversal invariant insulators, Phys. Rev. B **78**, 195424 (2008).

[25] N. Nagaosa, J. Sinova, S. Onoda, A. H. MacDonald, and N. P. Ong, Anomalous Hall effect, Rev. Mod. Phys. **82**, 1539 (2010).

[26] J. Wang, B. Lian, X.-L. Qi, and S.-C. Zhang, Quantized topological magnetoelectric effect of the zero-plateau quantum anomalous Hall state, Phys. Rev. B **92**, 081107 (2015).

[27] T. Morimoto, A. Furusaki, and N. Nagaosa, Topological magnetoelectric effects in thin films of topological insulators, Phys. Rev. B **92**, 085113 (2015).

[28] A. M. Essin, J. E. Moore, and D. Vanderbilt, Magnetoelectric polarizability and axion electrodynamics in crystalline insulators, Phys. Rev. Lett. **102**, 146805 (2009).

[29] S. S. Hong, J. J. Cha, D. Kong, and Y. Cui, Ultra-low carrier concentration and surface-dominant transport in antimony-doped $Bi_2Se_3$ topological insulator nanoribbons, Nat. Commun. **3**, 757 (2012).

[30] See supplementary material for further information about the sample growth, device dimensions, measurement setup, more detailed analyses of magnetotransport measurements, and other supporting data.

[31] S. Kasap and P. Capper, Springer Handbook of Electronic and Photonic Materials. Department of Electrical Engineering, University of Saskatchewan (2017). https://doi.org/10.1007/978-3-319-48933-9

[32] S. Mathimala, S. Sasmal, A. Bhardwaj, S. Abhaya, R. Pothala, S. Chaudhary, B. Satpati, and K. V. Raman, Signature of gate-controlled magnetism and localization effects at Bi$_2$Se$_3$/EuS interface, npj Quantum Mater. **5**, 64 (2020).

[33] J. Chen, H. J. Qin, F. Yang, J. Liu, T. Guan, F. M. Qu, G. H. Zhang, J. R. Shi, X. C. Xie C. L. Yang, K. H. Wu, Y. Q. Li, and L. Lu, Gate-voltage control of chemical potential and weak antilocalization in Bi$_2$Se$_3$, Phys. Rev. Lett. **105**, 176602 (2010).

[34] S. Hikami, A. I. Larkin, and Y. Nagaoka, Spin-orbit interaction and magnetoresistance in the two dimensional random system, Prog. Theor. Phys. **63**, 707 (1980).

[35] M. Liu, J. Zhang, C.-Z. Chang, Z. Zhang, X. Feng, K. Li, K. He, L.-L. Wang, X. Chen, X. Dai, Z. Fang, Q.-K. Xue, X. Ma, and Y. Wang, Crossover between weak antilocalization and weak Localization in a magnetically doped topological insulator, Phys. Rev. Lett. **108**, 036805 (2012).

[36] H.-T. He, G. Wang, T. Zhang, I.-K. Sou, G. K. L. Wong, J.-N. Wang, H.-Z. Lu, S.-Q. Shen, and F.-C. Zhang, Impurity effect on weak antilocalization in the topological insulator Bi$_2$Te$_3$, Phys. Rev. Lett. **106**, 166805 (2011).

[37] M. Liu, C.-Z. Chang, Z. Zhang, Y. Zhang, W. Ruan, K. He, L.-L. Wang, X. Chen, J.-F. Jia, S.-C. Zhang, Q.-K. Xue, X. Ma, and Y. Wang, Electron interaction-driven insulating ground state in Bi$_2$Se$_3$ topological insulators in the two-dimensional limit, Phys. Rev. B **83**, 165440 (2011).

[38] H.-Z. Lu and S.-Q. Shen, Weak localization of bulk channels in topological insulator thin films, Phys. Rev. B **84**, 125138 (2011).

[39] A. Schmid, Electron-phonon interaction in an impure metal, Z. Phys. **259**, 421 (1973).

[40] B. L. Altshuler, A. G. Aronov, and D. E. Khmelnitsky, Effects of electron-electron collisions with small energy transfers on quantum localisation, J. Phys. C: Solid State Phys. **15**, 7367 (1982).

[41] J. J. Lin and J. P. Bird, Recent experimental studies of electron dephasing in metal and semiconductor mesoscopic structures, J. Phys. Condens. Matter **14**, R501 (2002).

[42] P. A. Lee and T. V. Ramakrishnan, Disordered electronic systems, Rev. Mod. Phys. **57**, 287 (1985).

[43] L. N. Oveshnikov, V. A. Prudkoglyad, E. I. Nekhaeva, A. Y. Kuntsevich, Y. G. Selivanov, E. G. Chizhevskii, and B. A. Aronzon, Magnetotransport in thin epitaxial $Bi_2Se_3$ films, JETP Lett. **104**, 629 (2016).

[44] H.-Z. Lu and S.-Q. Shen, Weak localization and weak anti-localization in topological insulators, Proc. SPIE **9167**, 91672E (2014).

[45] Q. I. Yang, M. Dolev, L. Zhang, J. Zhao, A. D. Fried, E. Schemm, M. Liu, A. Palevski, A. F. Marshall, S. H. Risbud, and A. Kapitulnik, Emerging weak localization effects on a topological insulator–insulating ferromagnet ($Bi_2Se_3$-EuS) interface, Phys. Rev. B **88**, 081407(R) (2013).

[46] P. Wei, F. Katmis, B. A. Assaf, H. Steinberg, P. Jarillo-Herrero, D. Heiman, and J. S. Moodera, Exchange-coupling-induced symmetry breaking in topological insulators, Phys. Rev. Lett. **110**, 186807 (2013).

[47] F. Katmis, V. Lauter, F. S. Nogueira, B. A. Assaf, M. E. Jamer, P. Wei, B. Satpati, J. W. Freeland, I. Eremin, D. Heiman, P. Jarillo-Herrero, and J. S. Moodera, A high-temperature ferromagnetic topological insulating phase by proximity coupling, Nature **533**, 513 (2016).

# Supplementary Material

# Sb doping effect on transport behavior in the topological insulator $Bi_2Se_3$

Shu-Wei Wang[1,2,*], Hang Chi[3], and Jui-Che Chung[1,4]

[1]*Institute of Microelectronics, Department of Electrical Engineering, National Cheng Kung University, Tainan 701, Taiwan*
[2]*Academy of Innovative Semiconductor and Sustainable Manufacturing, National Cheng Kung University, Tainan 701, Taiwan*
[3]*Department of Physics, University of Ottawa, Ottawa, Ontario K1N 6N5, Canada*
[4]*Department of Physics, National Sun Yat-sen University, Kaohsiung 804, Taiwan*

*To whom correspondence should be addressed: shuweiwang@gs.ncku.edu.tw

This document contains:

1. Material growth and device fabrication
2. Measurement details
3. Hall effect analysis
4. Temperature dependence of the resistance ($R_{xx}$-$T$)
5. The phonon scattering effect on the mobility
6. The decoherence mechanism
7. Band structures of EuS/$Bi_{2-x}Sb_xSe_3$ heterostructures

# 1. Material growth and device fabrication

The growth of the $Bi_{2-x}Sb_xSe_3$ film was carried out in a homemade molecular beam epitaxy (MBE) system with a base pressure lower than $5\times10^{-10}$ mbar. The SrTiO$_3$ (111) substrates used for the growth of all the heterostructures were first soaked in 90 °C deionized water for 1.5 hours, and then annealed at 985 °C for 3 hours in a tube furnace with pure oxygen flow. Through the above steps, the surface of SrTiO$_3$ substrates were passivated and became atomically flat. These heat-treated SrTiO$_3$ substrates were then outgassed at ~ 530 °C for 1 hour before the growth of the $Bi_{2-x}Sb_xSe_3$ film. High purity Bi (99.999%), Sb (99.999%), and Se (99.999%) were co-evaporated from Knudsen effusion cells. The flux ratio of Se to Bi and Sb was set to be ~ 10:1. The growth rate for the films was controlled at ~ 0.2 nm per minute, and the SrTiO$_3$ substrates were maintained at 240 °C during the $Bi_{2-x}Sb_xSe_3$ film growth. The growth process was monitored *in situ* by reflection high energy electron diffraction (RHEED). After the growth, the $Bi_{2-x}Sb_xSe_3$ films were annealed at ~ 240 °C for 30 minutes to improve the crystal quality before being cooled down to room temperature. The sharp and streaky diffraction patterns in RHEED images indicate high crystal quality of the $Bi_{2-x}Sb_xSe_3$ films (Fig. S1).

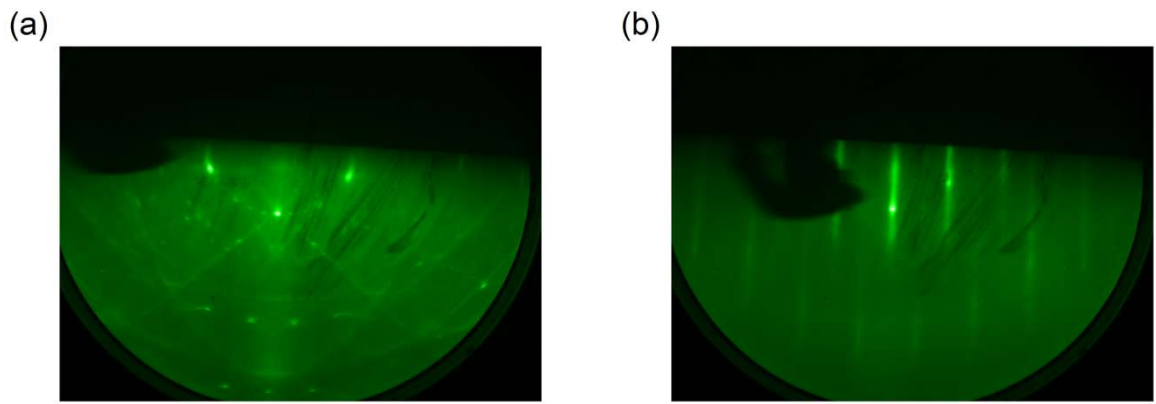

**Figure S1.** RHEED patterns of the substrate and the $Bi_{2-x}Sb_xSe_3$ film. (a) RHEED patterns of the heat-treated SrTiO$_3$ (111) substrate. (b) RHEED patterns of the $Bi_{2-x}Sb_xSe_3$ film. The shadow in the image is caused by heater wires between the electron source and the screen.

After the growth of the $Bi_{2-x}Sb_xSe_3$ film, a EuS layer of 3 nm was deposited *in situ* on the top of the $Bi_{2-x}Sb_xSe_3$ film at room temperature by an e-gun evaporator. The EuS layer is aimed for inducing ferromagnetic ordering in the adjacent $Bi_{2-x}Sb_xSe_3$ layer and opening an exchange gap in the topological surface band.

Contamination from the environment can substantially change the properties of TI films. Prior studies have shown that the size of the exchange gap is small and the environmental doping effect can easily shift the Fermi level out of the exchange gap [1-5], hindering the observation of the magnetic properties induced in the TIs. Therefore, a layer of 2-nm-thick amorphous $Al_2O_3$ was capped onto the EuS/$Bi_{2-x}Sb_xSe_3$ heterostructures before taking the sample out of the MBE chamber.

The samples were fabricated into Hall bars by standard e-beam lithography and ion milling dry etching. Then the samples were loaded into a thermal evaporator for metal deposition. Cr (~ 5 nm)/Au (~ 14 nm) were evaporated to form Ohmic contacts of the Hall bars. Finally, the Hall bar devices were mounted onto a leadless chip carrier (LCC) using GE varnish. The stack structure of the sample is schematically shown in Fig. S2(a). The dimension of the Hall bar is shown in Fig. S2(b). Figure S2(c) shows a typical optical image of the Hall bars. The dark part is the $SrTiO_3$ substrate. The light area is the EuS/$Bi_{2-x}Sb_xSe_3$ film. The yellow regions are the Cr/Au electrodes.

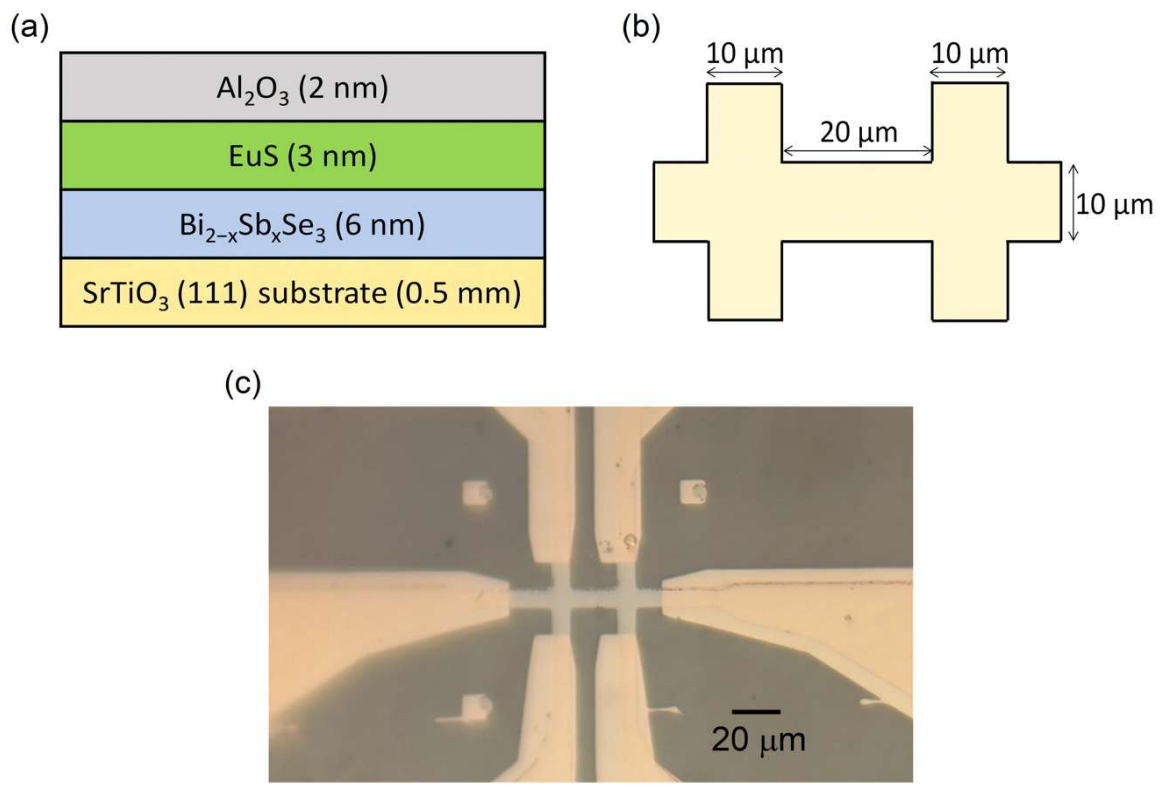

**Figure S2.** The stack structure and device geometry of the sample. (a) The layer structure of the EuS/$Bi_{2-x}Sb_xSe_3$ samples. (b) The dimensions of the Hall bar devices. (c) Typical optical image of the Hall bar device.

## 2. Measurement details

The transport measurements were carried out in a Quantum Design DynaCool Physical Property Measurement System (PPMS), which has a base temperature of 2 K and a superconducting magnet that can generate up to 9 T perpendicular magnetic field. The source-drain current ($I_{SD}$) used in the transport measurements is 10 μA. The longitudinal and Hall resistances are measured by the setup shown in Fig. S3.

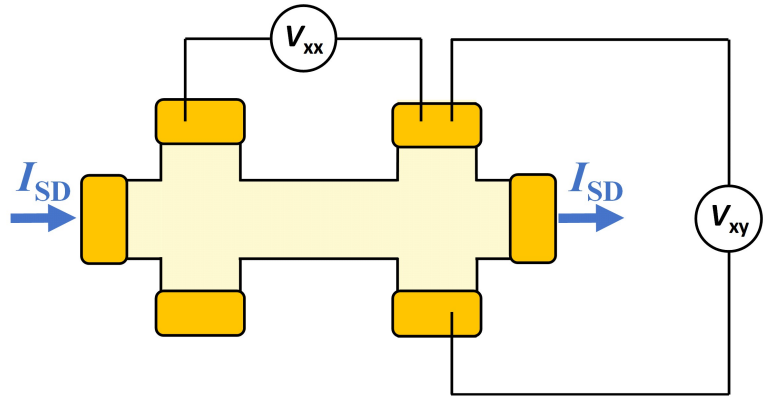

**Figure S3.** The transport measurement setup. The longitudinal resistance ($R_{xx}$) and the Hall resistance ($R_{xy}$) are obtained by $V_{xx}/I_{SD}$ and $V_{xy}/I_{SD}$, respectively.

## 3. Hall effect analysis

The Hall effect is observed in both EuS/Bi$_{2-x}$Sb$_x$Se$_3$ samples from 2 K to 60 K when an out-of-plane magnetic field is applied, as shown in Fig. S4. The negative sign of the Hall slope implies that the majority carriers in these two samples are both electrons. The Hall effect allows us to calculate the carrier concentrations by the relation $R_{xy}/\mu_0H = 1/ne$, where $R_{xy}$ is the Hall resistance, $\mu_0$ is the vacuum permeability, $H$ is the magnetic field strength, $n$ is the 2D carrier concentration, and $e$ is the elementary charge. After obtaining the carrier concentrations of the two EuS/Bi$_{2-x}$Sb$_x$Se$_3$ samples at different temperatures. The carrier mobility $\mu$ can be calculated by the relation $\sigma = ne\mu$. The temperature dependence of $\mu$ and $n$ are shown in Fig. S5. The carrier mobility decreases with the increasing temperature, indicating the electron-phonon scattering is the predominant scattering mechanism from 2 K to 60 K. In contrast, the carrier concentrations only have a slight fluctuation throughout this temperature range. The carrier concentrations obtained are $\sim 5 \times 10^{13}$ cm$^{-2}$ and $\sim 1 \times 10^{13}$ cm$^{-2}$ for the Bi$_{1.72}$Sb$_{0.28}$Se$_3$ and Bi$_{1.28}$Sb$_{0.72}$Se$_3$ films, respectively, which corroborates that the Fermi level is successfully tuned by Sb doping. The higher the Sb doping level, the closer the Fermi level is to the Dirac point of the topological surface band. The carrier concentration ($\sim 1 \times 10^{13}$ cm$^{-2}$) suggests that the Fermi level is well inside the bulk bandgap and close to the Dirac point of the surface band for the Bi$_{1.28}$Sb$_{0.72}$Se$_3$ sample.

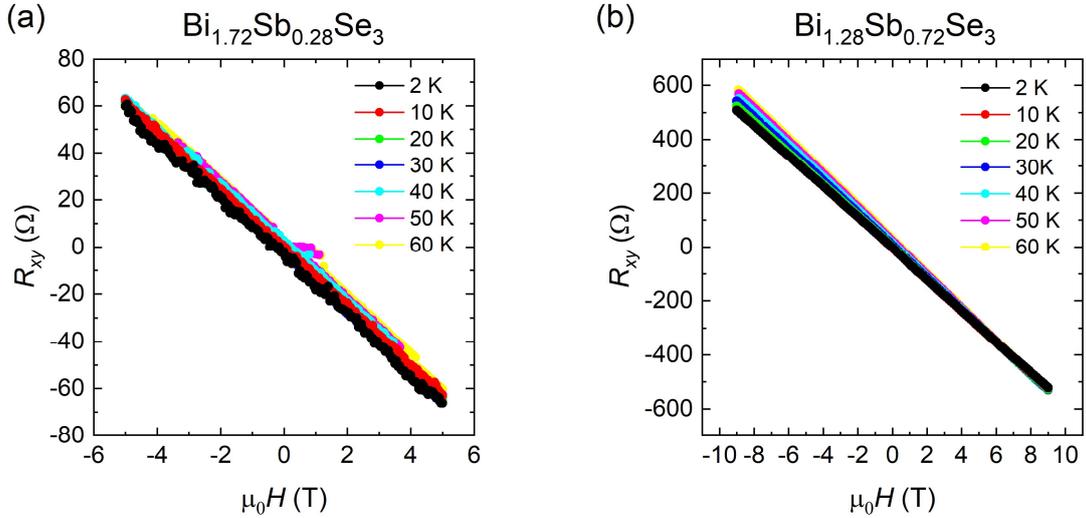

**Figure S4.** The Hall effect observed in EuS/Bi$_{2-x}$Sb$_x$Se$_3$ Hall bar devices. The Hall effect can be observed in both the Bi$_{1.72}$Sb$_{0.28}$Se$_3$ and Bi$_{1.28}$Sb$_{0.72}$Se$_3$ films between 2 K and 60 K when applying an out-of-plane magnetic field.

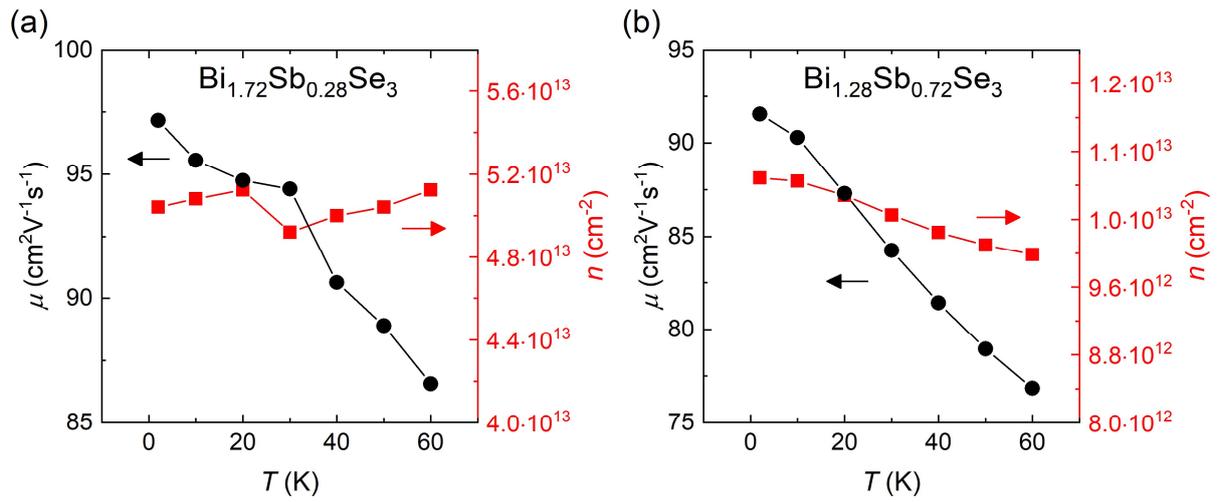

**Figure S5.** The temperature dependence of the carrier mobility and the carrier concentration. The carrier mobility (black dots) decreases as the temperature increases, while the carrier concentration (red squares) shows little change from 2 K to 60 K.

## 4. Temperature dependence of the resistance ($R_{xx}$-$T$)

The temperature ($T$) dependence of the longitudinal resistance ($R_{xx}$) is shown in Fig. S6. Both the $Bi_{1.72}Sb_{0.28}Se_3$ and $Bi_{1.28}Sb_{0.72}Se_3$ films show a metallic temperature dependence, i.e., the resistance decreases with the decreasing temperature, from room temperature to the base temperature (2 K). This suggests that the gapless surface states are dominant and the bulk states are suppressed throughout the entire temperature range. The observed metallic behaviour also agrees with previous reports with similar Sb doping concentrations in $Bi_2Se_3$ [6].

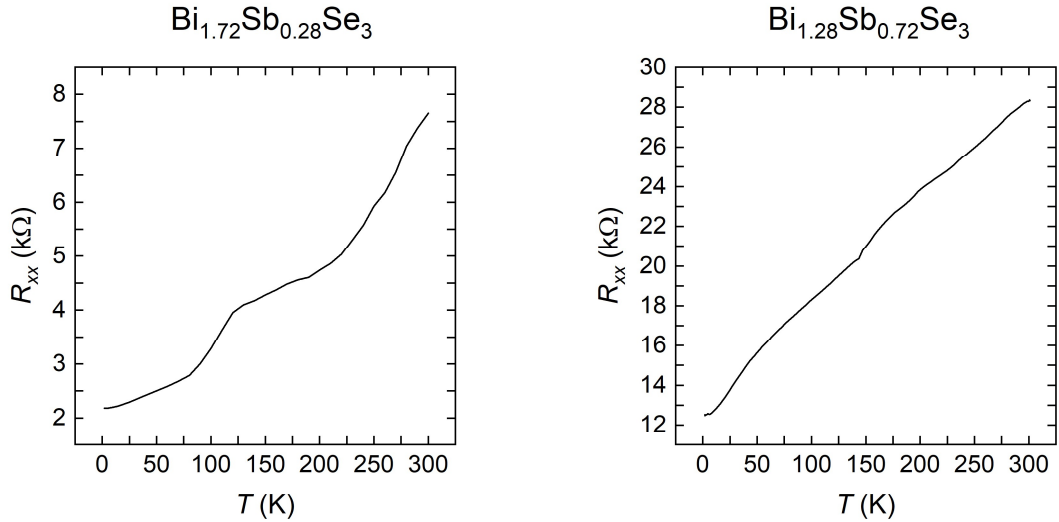

**Figure S6.** The temperature dependence of the longitudinal resistance. The temperature dependences of $R_{xx}$ of both $Bi_{1.72}Sb_{0.28}Se_3$ and $Bi_{1.28}Sb_{0.72}Se_3$ films show a metallic behavior from 2 K to 300 K.

## 5. The phonon scattering effect on the mobility

In the low temperature range in this study, phonon scattering and ionized impurity scattering are the two possible scattering mechanisms in the Sb-doped $Bi_2Se_3$ thin films. In the main text, we find that the carrier mobility $\mu$ decreases with the increasing temperature, which implies that the phonon scattering is the predominant scattering mechanism [7]. According to the phonon scattering model, the carrier mobility will decline with a $T^{3/2}$ dependence [7]. The linear fits in Fig. S7 clearly show that $\mu$ is inverse-proportional to $T^{3/2}$, corroborating the phonon-dominated scattering in both $Bi_{1.72}Sb_{0.28}Se_3$ and $Bi_{1.28}Sb_{0.72}Se_3$ films. We also calculated the standard error of the mobility data and obtained $\pm 1.4751$ cm$^2$/V·s and $\pm 2.1264$ cm$^2$/V·s for the $Bi_{1.72}Sb_{0.28}Se_3$ and $Bi_{1.28}Sb_{0.72}Se_3$ films, respectively. The value of the error is only ~1/10 of the variation range of the mobility, which suggests that the temperature dependence of the mobility is statistically significant.

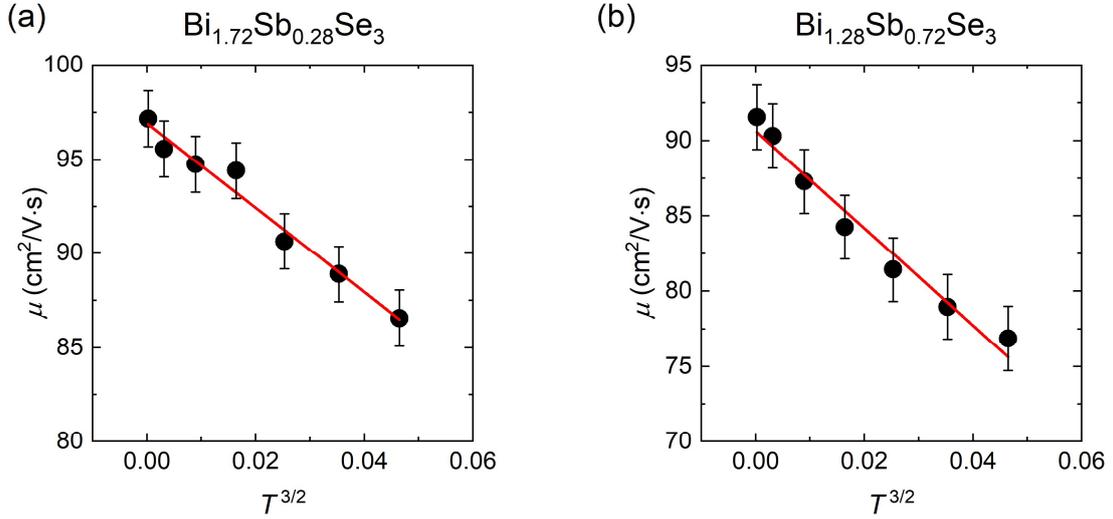

**Figure. S7.** The temperature dependence of the carrier mobility $\mu$. The left panel and the right panel show the mobility data (black dots with error bars) and the corresponding fitting lines (red lines) of the $Bi_{1.72}Sb_{0.28}Se_3$ and the $Bi_{1.28}Sb_{0.72}Se_3$ films, respectively. The fitting shows that $\mu$ is inversely proportional to $T^{3/2}$.

## 6. The decoherence mechanism

We observed that the predominant decoherence mechanism in both $Bi_{2-x}Sb_xSe_3$ films seems to be different in different temperature regimes. The phase coherence length in the $Bi_{1.72}Sb_{0.28}Se_3$ film is significantly larger than that in the $Bi_{1.28}Sb_{0.72}Se_3$ film below 30 K, but becomes comparable over 30 K. We speculate that, below 30 K, the electron-electron interaction caused by the doped impurities is not negligible and hence the decoherence mechanism is dominated by electron-electron interaction. Consequently, the phase coherence length is reduced more in the sample with heavier Sb doping and stronger impurity effect, i.e., the $Bi_{1.28}Sb_{0.72}Se_3$ film. When the temperature exceeds 30 K, the electron-electron interaction becomes less important and thermal phonon takes over the main role of the dephasing mechanism. Hence the phase coherence lengths in the two samples become similar above 30 K. To support our hypothesis, we calculated the standard error of the phase coherence length data and obtained the values of ±18.0271 nm and ±9.9332 nm for the $Bi_{1.72}Sb_{0.28}Se_3$ and $Bi_{1.28}Sb_{0.72}Se_3$ films, respectively. The coherence length data with error bar is shown in Fig. S8. Below 30 K, the coherence lengths for the $Bi_{1.72}Sb_{0.28}Se_3$ film are 150.3228 nm (2 K), 108.0266 nm (10 K) and 66.4117 nm (20 K). On the other hand, the coherence lengths for the $Bi_{1.28}Sb_{0.72}Se_3$ film below 30 K are 86.3479 nm (2 K), 71.2811 nm (10 K), and 55.6117 nm (20 K). Taking into account the calculated error, the coherence length in $Bi_{1.72}Sb_{0.28}Se_3$ is consistently larger than that in $Bi_{1.28}Sb_{0.72}Se_3$ below 30 K, indicating a statistically meaningful difference even with the uncertainties. Above 30 K, the coherence length data with overlapping error bars indicate that the $l_\phi$ in the two samples becomes comparable within error margins above 30 K. This result supports our speculation regarding the shift of the predominant dephasing mechanism in different temperature regimes.

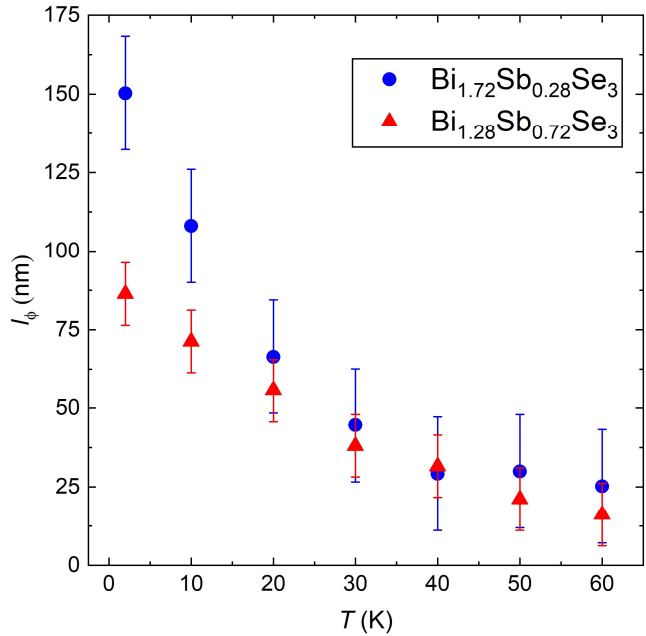

**Figure. S8.** The phase coherence length as a function of the temperature. Blue dots and red triangles are the $T$ dependence of $l_\phi$ in $Bi_{1.72}Sb_{0.28}Se_3$ and $Bi_{1.28}Sb_{0.72}Se_3$ films, respectively.

## 7. Band structures of EuS/Bi$_{2-x}$Sb$_x$Se$_3$ heterostructures

The observed WAL suggests that the Fermi level is well inside the bulk bandgap and crosses the topological surface band for both Bi$_{1.72}$Sb$_{0.28}$Se$_3$ and Bi$_{1.28}$Sb$_{0.72}$Se$_3$ films. In the Hall measurements (Fig. S4), both Bi$_{1.72}$Sb$_{0.28}$Se$_3$ and Bi$_{1.28}$Sb$_{0.72}$Se$_3$ films exhibit *n*-type transport, which implies that the Fermi levels crosses the conduction band of the topological surface states. The carrier concentration in the Bi$_{1.28}$Sb$_{0.72}$Se$_3$ film is much lower than that in the Bi$_{1.72}$Sb$_{0.28}$Se$_3$ film, meaning that the doping of Sb atoms successfully changes the position of the Fermi level and moves it closer to the Dirac point. Finally, as described in the main text, the EuS layer is not exchange coupled to the TI layer, hence the magnetic gap is not opened in the surface band of both Bi$_{1.72}$Sb$_{0.28}$Se$_3$ and Bi$_{1.28}$Sb$_{0.72}$Se$_3$. With all the information above, the band structures of the EuS/Bi$_{1.72}$Sb$_{0.28}$Se$_3$ and EuS/Bi$_{1.28}$Sb$_{0.72}$Se$_3$ samples can be inferred to be like that shown in Fig. S9.

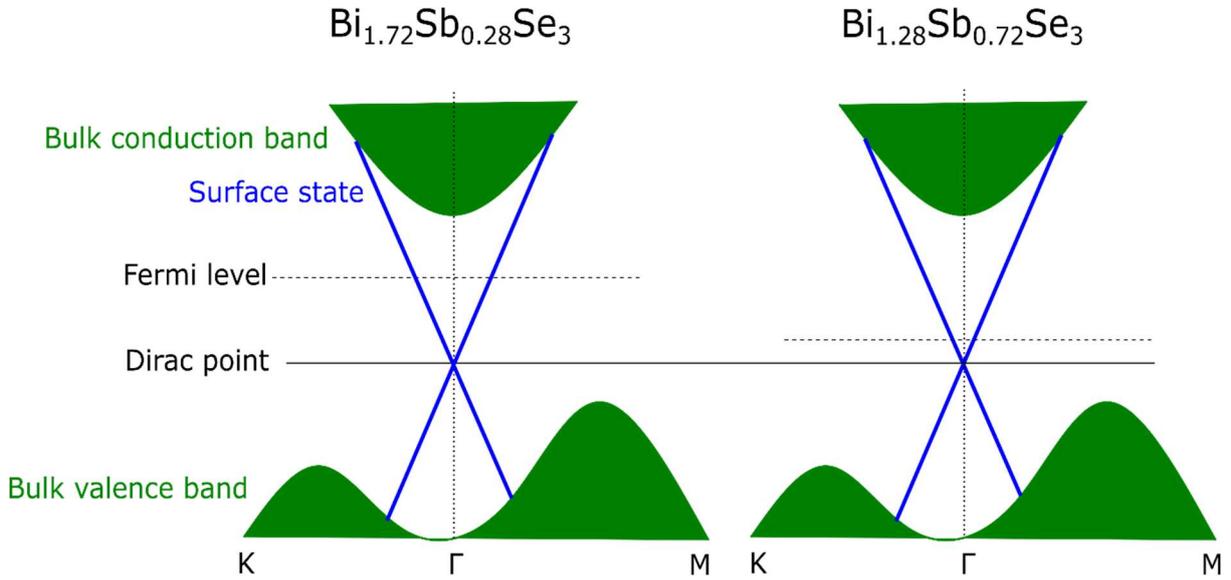

**Figure. S9.** The inferred band structures of the EuS/Bi$_{2-x}$Sb$_x$Se$_3$ heterostructures. The left panel and the right panel show the band structures of the EuS/Bi$_{1.72}$Sb$_{0.28}$Se$_3$ and the EuS/Bi$_{1.28}$Sb$_{0.72}$Se$_3$ films, respectively. The green parts are the bulk bands. The blue lines depict the topological surface bands. The two dotted lines indicate the positions of the Fermi levels in the two films, respectively, which are different due to the different Sb doping concentrations. The magnetic gap is not opened in both films.

# References

[1] A. J. Bestwick, E. J. Fox, X. Kou, L. Pan, K. L. Wang, and D. Goldhaber-Gordon, Phys. Rev. Lett. **114**, 187201 (2015).

[2] C.-Z. Chang, W. Zhao, D. Y. Kim, P. Wei, J. K. Jain, C. Liu, M. H. W. Chan, and J. S. Moodera, Phys. Rev. Lett. **115**, 057206 (2015).

[3] M. Liu, W. Wang, A. R. Richardella, A. Kandala, J. Li, A. Yazdani, N. Samarth, and N. P. Ong, Sci. Adv. **2**, 7 (2016).

[4] M. Kawamura, R. Yoshimi, A. Tsukazaki, K. S. Takahashi, M. Kawasaki, and Y. Tokura, Phys. Rev. Lett. **119**, 016803 (2017).

[5] E. J. Fox, I. T. Rosen, Y. Yang, G. R. Jones, R. E. Elmquist, X. Kou, L. Pan, K. L. Wang, and D. Goldhaber-Gordon, Phys. Rev. B **98**, 075145 (2018).

[6] J. Zhang, C.-Z. Chang, Z. Zhang, J. Wen, X. Feng, K. Li, M. Liu, K. He, L. Wang, X. Chen, Q.-K. Xue, X. Ma and Y. Wang, Nat. Commun. **2**, 574 (2011).

[7] S. Kasap and P. Capper, Springer Handbook of Electronic and Photonic Materials. Department of Electrical Engineering, University of Saskatchewan (2017).